\documentclass[letterpaper]{article}
\usepackage{aaai}
\usepackage{times} 
\usepackage{helvet} 
\usepackage{courier} 
\usepackage{verbatim}
\usepackage{amsmath}
\usepackage{amssymb}
\usepackage{url}
\usepackage{array}
\usepackage{booktabs}
\usepackage{afterpage}
\usepackage[dvipsnames]{color}
\usepackage{mfirstuc}
\usepackage{subfigure}
\makeatletter
\newif\if@restonecol
\makeatother

\usepackage[lined,algonl,boxed]{algorithm2e}\usepackage{epsfig}
\usepackage{multirow}
\usepackage{rotating}
\title{
On the Interplay between Social and
Topical Structure\thanks{All authors contributed equally to this work.}
}

\author{{Daniel M. Romero}\\
{Northwestern Institute on Complex Systems}\\
{Northwestern University}\\
{Evanston, IL}\\
d-romero@kellogg.northwestern.edu
\And
 Chenhao Tan\\
 {Dept. of Computer Science}\\
 {Cornell University}\\
 {Ithaca, NY}\\
 chenhao@cs.cornell.edu
 \And
 Johan Ugander \\
 {Center for Applied Mathematics}\\
 {Cornell University}\\
 {Ithaca, NY}\\
 jhu5@cornell.edu
}
       
\begin{document}
\maketitle
\vfil\eject
\begin{abstract}

People's interests and people's social relationships are intuitively
connected, but understanding their interplay and 
whether they can help predict each other has remained an open question.
We examine the interface of two decisive structures forming the
backbone of online social media: the graph structure of social
networks --- who 
connects with whom --- and the set structure of
topical affiliations --- who is interested in what.
In studying this interface, we identify key relationships whereby each of these
structures can be understood in terms of the other. 
The context for our analysis is Twitter, a complex 
social network of both follower relationships and
communication relationships.
On Twitter, ``hashtags'' are used to label conversation topics, 
 and we examine hashtag usage alongside
these social structures.
 
We find that the hashtags that users adopt can predict
their social relationships, and also that the social
relationships between the initial adopters of a hashtag 
can predict the future popularity of that hashtag. 
By studying weighted social relationships, 
we observe that while strong reciprocated ties are 
the easiest to predict from hashtag structure, 
they are also much less useful than weak directed ties
for predicting hashtag popularity.
Importantly, we show that 
computationally simple structural determinants 
can provide remarkable performance in both tasks.
While our analyses focus on Twitter, we view our findings  as broadly applicable to topical
affiliations and social relationships in a host of diverse contexts, including the movies people watch, the brands people like, or the locations people frequent.
\end{abstract}

\section{Introduction}

Online social networks and online information sharing have gained tremendous popularity over the past several years, providing enormous opportunities to track both how relationships form and how information diffuses in an online setting.
Two intertwined questions of fundamental interest are
{\em whether 
the network of social relationships can 
predict the popularity of topics of interest
 among users} 
and 
{\em whether the topics of interests
can predict the network of relationships between users}.
In this work, we aim to bring these relationships and 
interests together and study the ways in which they are related.
To do this, we investigate the extent to which the relation between social and 
topical aspects can be used to predict each other with 
basic, interpretable features, and guide the investigation
with basic predictive modeling using logistic regression.
Concretely, we examine 
two key structures of online social media -- the set structure of 
topical affiliations and the graph structure of social networks. 

For topical affiliations, we consider a user in a social network to be tagged with a certain
topic if information related to that topic passes through the user, 
or if there is evidence that the user is interested in the topic. 
For example, when a Twitter user posts a tweet with a certain hashtag, 
we affiliate them with that hashtag. 
In other contexts providing digital traces of topical affiliations, topics may correspond to musical artists, movies, or restaurants. 
From the users affiliated with each topic we derive a set, and 
examining all topics generates a 
set structure of topical affiliations. 
In this work, we show that this set structure can be useful in
understanding the structure and evolution of the network itself.

First, we ask {\em how can we understand social structure in terms of topical structure?}
This amounts to a classic problem of link prediction in social networks, 
but with a focus on using only features from the set structure of topical affiliations. 
Many people have studied this problem and the approach has often been to 
look at the features of the existing network in order to predict future 
connections \cite{Liben-Nowell:2007,Taskar03linkprediction,Schifanella:2010,Leroy:2010,Rossetti:2011,Hutto+etal:2013}. 
In this work, we base our predictions exclusively on features related 
to the topical proximity of the users as well as the graph properties of the 
induced subgraph among the users who are tagged with these topics.
This part of the paper shows how the topical structure of the users in the 
network can inform our understanding of the structure of the network itself. 

Next, we ask {\em how can we understand topical structure in terms of social structure?}
Here we study a novel problem. We try to understand the extent to which the structure 
of the network determines the topical information diffusion process. 
In particular, we are interested in whether the structure of the graph induced by the initial 
set of adopters of a certain topic can tell us something about the eventual popularity of the topic.
Here, the word ``topic'' can be interpreted in a loose sense, referring to a product, 
an idea, or even a behavior. The idea that the speed and magnitude of adoption of products, 
ideas, and behaviors can be driven by ``viral marketing'' techniques has gained tremendous
 popularity over the past few years
 \cite{Brown:1987,Goldenberg:2001,Richardson:2002,Leskovec:2007,Rogers:1995,Gruhl:2004,Kimura:2006,Tsur+Rappoport:2012}. 
The premise is that one can utilize the edges of an existing social network
as bridges for information to spread from person to person. 
A common challenge is how to maximize the spread of a topic.
The focus of our study is to ask: how is this growth related to nascent graph structure?
To shed light on this question, we test a predictive algorithm that takes properties
of the induced subgraph of the early adopters of a topic as features, 
with the goal of predicting its eventual popularity.

Our particular domain of study is Twitter. We use \emph{hashtags} --- 
labels that users include in their posts to indicate the topic of the message --- 
to distinguish between different topics.
As for social relationships, we consider both the \emph{follower} and \emph{@-message}
communication networks as proxies for social connections among the users. 
Because these networks are directed and the @-message network is weighted, 
we are able to compare the differences in our results when considering strong
and weak ties, as well as reciprocated and unreciprocated relationships.

\noindent\textbf{Contributions.} 
First, we demonstrate how merely basic features from topical affiliations can 
yield remarkable prediction accuracy in link prediction. 
The performance can be further improved by incorporating social graph 
properties of these topics. 
Second, we find that the structure of the subgraphs induced by 
early adopters can indeed have predictive power about the eventual 
popularity of the topic. Furthermore, we observe that the relationship 
between the topological properties of the initial graphs and the topic's 
popularity is not always straight forward. For example, popularity of a topic 
does not monotonically change with the number of social connections 
among the initial users. Instead, we find that the topic gains 
popularity specifically when the number of connections is either very high or very low. 
This could come as a surprise since few connections among initial adopters of 
a topic could be perceived as the topic lacking ``virality.'' 
Bridging these two findings, we discover that while strong reciprocated ties are 
the easiest to predict from hashtag structure, they are also much less useful
than weak directed ties for predicting 
hashtag popularity. Our hypotheses for possibly explaining these 
observations are also discussed. 

Throughout this work we use naive baselines to compare with our prediction results.
The reason for that is that we do not aim to present sophisticated algorithms and techniques
that explicitly maximize the accuracy of link and popularity prediction.
We instead aim to expose the relationship that topical structure and social
structure share by showing that \emph{simple} features of topical affiliations 
can predict social relationships and that \emph{simple} 
features of social structure can predict topical popularity, 
even through standard predictive models.

%\vspace{-0.05in}

\section{Dataset}
The dataset used in this paper consists of two main parts: hashtags and networks.
During January 2010, Twitter was crawled using their publicly available API.
The last 3,200 tweets of each existing user were collected. Over 95\% of the users in this dataset have less than 3,200 tweets. Hence, the dataset contains a complete history of tweets for almost all users. 
Overall, over three billion messages from more than 60 million users
were obtained during this crawl \cite{romero:www2011}. \\

\vspace{-0.05in}

\noindent{\bf Hashtags.} A convention widely used by Twitter users is a tagging system where a user includes a single undelimited string proceeded by a ``\#'' character. This string is termed a \emph{hashtag}, and it is meant to label the tweet in order to signal to others that it belongs to a particular conversation topic. For example: ``\emph{What a thrilling game last night between the Thunder and Grizzlies. \#NBA.''} From our dataset we extracted a total of 7,305,414 hashtags, with 5,513,587 users who utilized at least one hashtag. On average, each hashtag was used by 9.48 distinct users, while a user posted about 12.57 different hashtags. \\

\vspace{-0.05in}

\noindent{\bf Graphs.} We obtained the follower/followee graph from \cite{Kwak:www10}, which contained the list of people each person was following at crawl time. If user $A$ follows user $B$ we create the edge $(A,B)$. There are around 366 million edges among the users who have utilized at least one hashtag.
The second graph is based on \emph{@-messages}. These are posts that are publicly directed from one user to another. They can be used as directed messages, or to reference another user. To send an @-message, the sending user includes the ``@'' character next to the receiving person's username in their tweet. 
The @-graph is created by constructing the edge $(A,B)$ if $A$ has sent at least $n$ @-messages to $B$ in the tweets available ($n$ is a threshold to indicate the strength of the relationship between two users). There are around 85 million edges in the @-graph with threshold $n=1$ among the users who have utilized at least one hashtag.

\section{Social Structure from Topical Structure}

In this section, we investigate the social structure from the topical 
affiliation system
 defined by the usage of hashtags.
Specifically, we ask to what extent a user's hashtags reveal their ties to other users in Twitter's directed follower graph, or their ties to other users via $@$-messages. By allowing hashtags to define user sets, we can view Twitter users as embedded in the topical affiliation system of these hashtags. 
We focus on the features extracted from the topical affiliations in the hope that this topical affiliation system helps us understand the social structure.
We consider the 
link prediction problem, where we are trying to predict the presence of an edge between arbitrary pairs of individuals. From this, we observe that the popularity of the least popular common hashtag that two users overlap on is a surprisingly informative predictor. 
Having observed this, we demonstrate that adding simple features of the graph structure of these least popular common hashtags can be used to  develop a capable predictive model. 

\noindent{\bf Measuring topic distance.} In order to approach this question, we must first summarize the hashtag usage similarity of two individuals from the topical affiliation system into features that could plausibly serve as similarity/distance measures.  Perhaps the most obvious measure of similarity is the number of sets that contain two users. In our context this corresponds to the number of hashtags that two users have in common.  This measure is immediately problematic, since it does not distinguish between hashtags that are broadly adopted and those that have only been used by a handful of users. To address this issue, we define \emph{the size of a hashtag} to be the number of people who have used the hashtag, and consider features that relate to the size of the common hashtags. For this, we consider the size of the smallest common hashtag, the size of the largest common hashtag, the average size, and also two measures that aggregate the common overlap of the full sets: the sum distance and the Adamic-Adar distance \cite{adamic:2003}. 

Consider the following notation:
\begin{itemize}
\item Let $u_1, \ldots, u_N$ be the $N$ users.
\vspace{-0.04in}
\item Let $h_1, \ldots, h_M$ be the $M$ hashtags.
\vspace{-0.04in}
\item Let $H(u_i)$ be the set of hashtags used by users $u_i$.
\vspace{-0.04in}
\item Let $U(h_j)$ be the users who used hashtag $h_j$.
\end{itemize}

The features of the common hashtags between users that we consider in the work are:

\begin{itemize}
\item The number of hashtags in common,\\ $| H(u_i) \cap H(u_j) |$.
\vspace{-0.04in}
\item The size of the smallest common hashtag,\\ $\min_{h \in H(u) \cap H(v)} |U(h)|$.
\vspace{-0.04in}
\item The size of the largest common hashtag,\\ $\max_{h \in H(u) \cap H(v)} |U(h)|$.
\vspace{-0.04in}
\item The average size of the common hashtags,\\ $\frac{1}{|H(u) \cap H(v)|} \sum_{h \in H(u) \cap H(v)} |U(h)|$.
\vspace{-0.04in}
\item The sum of the inverse sizes,\\ $\sum_{h \in H(u) \cap H(v)} 1/|U(h)|$.
\vspace{-0.04in}
\item The Adamic-Adar distance,\\  $\sum_{h \in H(u) \cap H(v)} 1/\log|U(h)| $.
\end{itemize}

\begin{figure}[thb!]
\centering
\includegraphics[width = 0.48\textwidth]{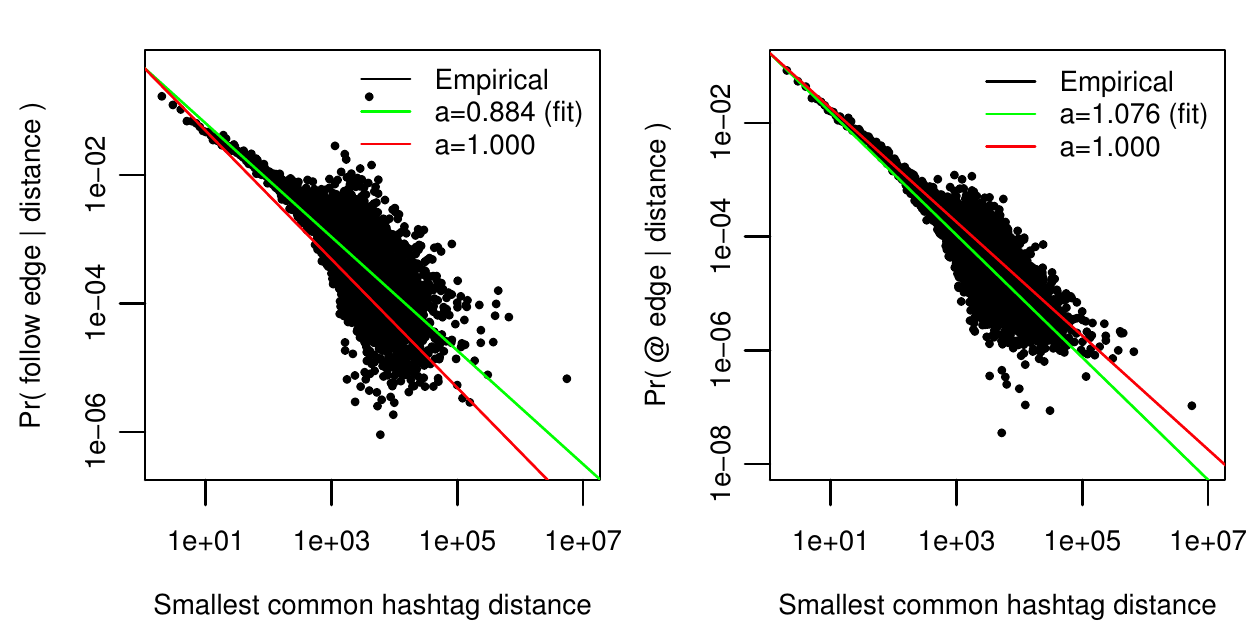} 
\caption{Linkage probability as a function of smallest common hashtag. (a) The probability of a given user following another user as a function of the size, and (b) the probability of a given user @-messaging another user as a function of size. Both figures are shown on a log-log scale. \label{PrLink}}
\end{figure}

The size of the smallest common hashtag is an intuitively attractive measure 
since it captures the extent to which the conversations two persons share are 
unique or not. It also happens to have a significance in the study of 
decentralized information routing.
In \cite{kleinberg:2002}, Kleinberg studied social networks where individuals were viewed as embedded in a set system (much like this hashtag topical affiliation system) and an individual $u$ was linked to an individual $v$ with probability proportional to $d(u,v)^{-a}$, where $d(u,v)$ is the size of the smallest common set. Kleinberg showed that decentralized greedy routing with regard to this measure takes polylogarithmic time if and only if $a = 1$. We observe that this dependence between smallest common size and link probability
approximately holds true in our setting.
From Figure \ref{PrLink} we see that the probability of a link as a function 
of the smallest common hashtag size appears to obey an inverse power law. 
In the case of @-communication (here thresholded on 3 @-messages),
the probability of linkage appears to be quite close to $a=1$, 
the condition for efficient decentralized greedy routing.

The size of the largest common hashtag is an intuitively poor feature for predicting links, and we include it here specifically as a 
dummy feature, showing that not all features of the common hashtags are informative. Very commonly, the largest common hashtag that users overlap on is one of a few extremely popular hashtags, for example, \#musicmonday, \#ff, or \#fail.
In choosing to study the sum distance and Adamic-Adar distance --- 
a measure introduced for studying the similarity of web-based 
social networks derived from common homepage content --- we allow ourselves to consider similarity measures using all common hashtags (topics). \\

\noindent{\bf Predictive model.} Given the features defined above, we now investigate how well the topical overlap represented in the common hashtags allows us to predict the presence of links. We formulate this task as a balanced classification task: given a random set of 100,000 user pairs that coincide on at least one hashtag,\footnote{We consider only user pairs that coincide on at least one hashtag because performing classification against completely arbitrary user pairs (where only $4.9\%$ of hashtag-using user pairs coincide on some hashtag) would have been overly generous to our classifier. Arbitrary users rarely coincide on any hashtags, and the coincidence rates between connected users is understandably going to be higher. In fact, $35\%$ of user pairs where one user follows the other coincide on some hashtag, and fully $78\%$ coincide when comparing user pairs where one person has @-messaged the other person at least once. 
}
where 50,000 are disconnected pairs and 50,000 are connected, what sort of prediction accuracy can we obtain, compared to a naive 50\% baseline? 

We approach the problem using logistic regression with 10-fold cross-validation. For all six features, in addition to a linear term, we also include the logarithm and the inverse of each value, allowing us to more robustly extract non-linear dependencies. 

We consider the tasks of predicting follow edges, mutual follow edges, @-message edges, and mutual @-message edges.
It is natural to view the number of @-messages as the strength of a tie, and we therefore also consider our classification task applied to @-message edges thresholded on different @-message counts. Note that for each of the tasks, we use a sample of 100,000 user pairs that coincide on at least one hashtag where half of the pairs are connected and half disconnected according to the particular definition of edges in the task.\footnote{Note that we have different sets of user pairs for different prediction tasks to keep the dataset balanced.}  Therefore the naive baseline is 50\% for all the different tasks. The accuracies we obtain are shown in Table~\ref{t:links}, where we report the performance of both a full model, considering all features under all transformations, and also models trained on a single feature set (where we include its two transformations). 

\begin{table*}
\scriptsize
\centering
\begin{tabular}{cc}
\raisebox{-3em}{
\begin{sideways}
Directed edges
\end{sideways}} &
\begin{tabular}{c||c|c|c|c|c|c|c}
Model Features & Follow & @ $\ge1$ & @ $\ge3$ &@ $\ge5$ & @ $\ge7$ &  @ $\ge9$ & @ $\ge20$\\ 
\hline
\hline
All hashtag features & 0.737 & 0.826 & 0.850 & 0.860 & 0.862 & 0.870 & 0.871 \\
\hline
\# common HTs & 0.713 & 0.781 & 0.798 & 0.800 & 0.804 & 0.809 & 0.816 \\
Smallest HT size & 0.703 & 0.799 & 0.828 & 0.841 & 0.842 & 0.854 & 0.855 \\
Largest HT size & 0.582  & 0.587 & 0.584 & 0.585 & 0.581 & 0.583 &  0.575 \\
Average HT size & 0.589  & 0.662 & 0.683 & 0.702 & 0.697 & 0.723 & 0.720 \\
Sum distance & 0.712 & 0.804 & 0.832 & 0.845 & 0.848 & 0.858 & 0.860 \\
Adamic-Adar distance & 0.727 & 0.809 & 0.831 & 0.842 & 0.846 & 0.852 & 0.856 \\
\hline
\hline
Hashtag features + Edges & \bf 0.766 & \bf 0.863 & \bf 0.889 & \bf 0.921 & \bf 0.940 & \bf 0.949 & \bf 0.976\\
Edges of smallest & 0.647 & 0.790 & 0.816 & 0.827 & 0.865 & 0.872 & 0.886\\
\hline
\end{tabular}
\end{tabular}

\vspace{0.15in}
\begin{tabular}{cc}
\raisebox{-3em}{
\begin{sideways}
Mutual edges
\end{sideways}} &
\begin{tabular}{c||c|c|c|c|c|c|c}
Model Features &  Follow & @ $\ge1$ & @ $\ge3$ & @ $\ge5$ &  @ $\ge7$ &  @ $\ge9$ &  @ $\ge20$ \\ 
\hline
\hline
All hashtag features & 0.762 & 0.827 & 0.868 & 0.869 & 0.868 & 0.867 & 0.866 \\
\hline
\# common HTs & 0.739 & 0.782 & 0.809 & 0.813 & 0.812 & 0.812 & 0.808\\
Smallest HT size & 0.715 & 0.803 & 0.849 & 0.853 & 0.852 & 0.852 & 0.856 \\
Largest HT size & 0.576 & 0.562 & 0.590 & 0.583 & 0.574 & 0.569 & 0.548 \\
Average HT size & 0.597 & 0.671 & 0.712 & 0.706 & 0.707 & 0.706 & 0.743 \\
Sum distance & 0.725 & 0.808 & 0.854 & 0.857 & 0.856 & 0.856 & 0.860 \\
Adamic-Adar distance & 0.751 & 0.807 & 0.850 & 0.854 & 0.852 & 0.852 & 0.849 \\
\hline
\hline
Hashtag features + Edges & \bf 0.796 & \bf 0.864  & \bf 0.922 & \bf 0.936 & \bf 0.934 & \bf 0.949 & \bf 0.967\\
Edges of smallest & 0.651 & 0.788 & 0.829 &0.832 & 0.833 & 0.837 & 0.861 \\
\hline
\end{tabular}
\end{tabular}
\vspace{0.1in}
\caption{Prediction accuracies for directed and mutual edges, as trained on the full set of hashtag features, individual hashtag features, and edge features. Accuracy was evaluated using 10-fold cross-validation on a balanced classification dataset.\label{t:links}}
\vspace{0.05in}
\normalsize
\end{table*}

We see that classification based on common hashtag usage exhibits powerful prediction accuracy given its simplicity: for follow edges our accuracy is 74\%, for @-edges it is 83\%, and for strong @-edges (more than 20 messages), our accuracy is 87\%. We achieve comparable performance when predicting mutual edge relationships. Beyond this general performance, we note that the size of the smallest common hashtag is a consistently accurate feature when considered alone, especially when trying to predict strong ties. As expected, the size of the largest common hashtag is least informative.\\

\noindent{\bf Predictive model with edges.} The features we consider above do not extract anything about the graph structure of the induced subgraphs that each hashtag defines.\footnote{The induced subgraph from a hashtag refers to the edges between the users who have used that hashtag.} 
Extending the high accuracy of models based solely 
on the size of the smallest common hashtag, we now show
that a simple feature of the subgraph of the smallest common
hashtags -- its edge density -- can further improve our predictions. 

\begin{figure}[htb!]
\vspace{-0.1in}
\centering
\includegraphics[width =.48 \textwidth]{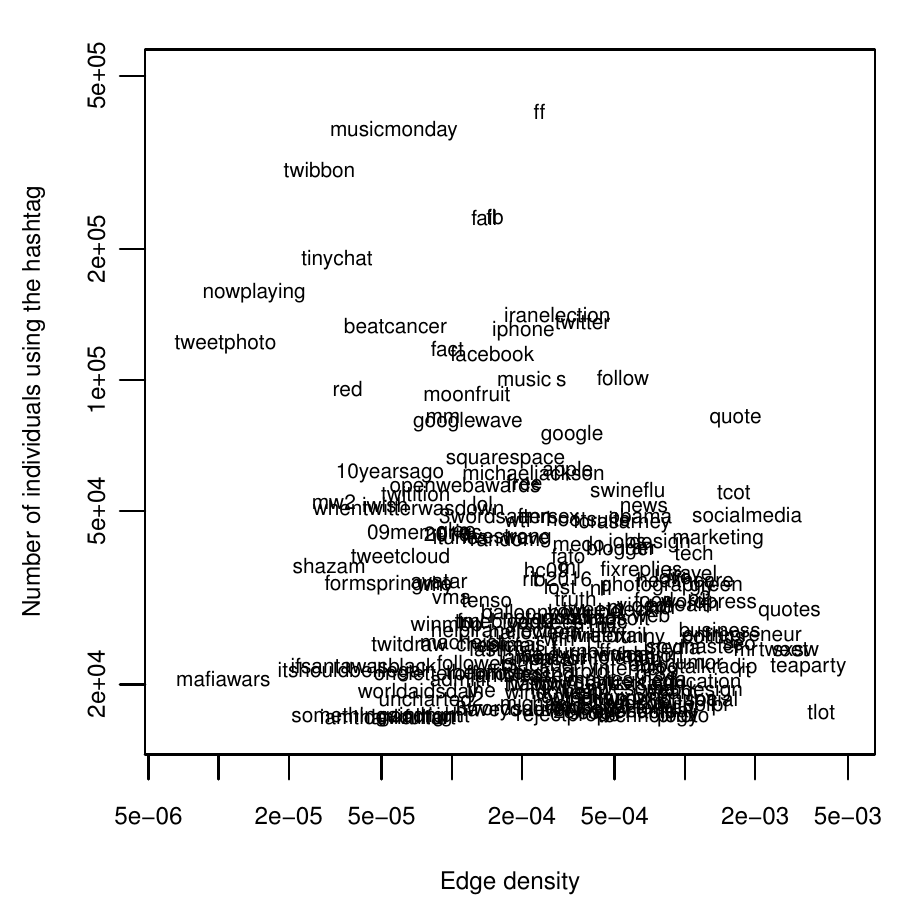} 
\vspace{-0.2in}
\caption{Edge density heterogeneity for the 200 most common hashtags in the dataset. \label{words}}
\end{figure}

One motivation for considering the graph structure is the observation, shown in Figure~\ref{words}, that the edge density for similarly sized hashtags can differ considerably. 
Edge density for a hashtag is defined as the probability that two users, who have utilized the hashtag, are connected.\footnote{Formally, edge density for a hashtag is defined as $\frac{|\{(u,v) \in E | u \in U(h), v \in U(h) \}|}{|U(h)|(|U(h)|-1)}$. $E$ can be either the set of edges in the follower/followee graph or in the @-graph. } Figure~\ref{words} shows the density of each hashtag in the follower graph as a function of the size of the hashtag. Using the @-graph produces a similar figure. 

Simply put, some hashtags --- some topics --- are much more ``social'' than others. For example, \emph{\#teaparty} and \emph{\#mafiawars} have similar popularity, however, the users in \emph{\#teaparty} are much more connected to each other, a plausible indicator that \emph{\#teaparty} is more ``social.'' In another context, where instead of hashtags we considered geographic locations as topics, this would translate to knowing that two people both visit a bar versus knowing that they both visit the same bank. Both may be equally unpopular locations to visit, but the bar is a decidedly more social environment, and two people having visited the same bar is clearly going to predict more social interaction than having visited the same bank. 

We would like to include the count of edges appearing in the smallest common hashtag as a feature in our prediction tasks. However, it is important to avoid inadvertently incorporating a circular reference whereby the link we are trying to predict is directly present in this feature. For example, consider the problem of predicting an edge between a pair of users where the induced subgraph for their smallest common hashtag has size two and one edge. If we don't make any correction, we would know for certain that these two users are connected. We therefore let our edge count feature be the number of edges present in the smallest common hashtag between users {\it other} than the two users being considered.

By including the count of such edges appearing in the smallest common hashtag as a feature, where the type of the edges is the same as the type we are trying to predict, we see that our classification performance becomes more accurate, demonstrating an accuracy of  76.6\% when classifying follower relationships and 97.6\% when classifying strongly tied @-edges. This improvement is consistent across all the predictions tasks we consider.

\section{Topical Structure from Social Structure}

As we observed in the previous section, the topical affiliation structure of a social network is related to its social structure and can be used to predict social links. In this section, we aim to exploit 
the topical structure from 
the characteristics of the network.
Specifically in Twitter, we try to predict the future popularity of hashtags with simple features of the graph induced by the initial adopters of hashtags. \\

 \noindent{\bf Correlations between initial graph structure and popularity.} The properties of the graph of initial adopters of a hashtag can suggest properties of the diffusion mechanism of the hashtag. For example, if the graph has a very large number of edges in it, a hypothesis could be that the adopters found out about the hashtag from each other, and that many of their friends who haven't adopted it yet are likely to use it in the future. The hashtag would then become popular ``virally''. On the other hand, it could be that the hashtag describes a topic in a small coherent community and that its growth is limited by the size of this community. If there are very few edges in the graph, a hypothesis could be that the initial adopters did not discover the hashtag through their connections, since their connections had not adopted it yet, which means that users are not ``virally'' adopting the hashtag and hence it will not become popular. On the other hand, it is possible that this hashtag will soon become very popular because it corresponds to an important exogenous event (outside of Twitter) and people are using it without ``discovering'' it through their connections. More generally, these hypotheses expose the question of whether the eventual popularity of a topic depends on the number of edges in the initial graph, and if so, whether high popularity of the topic is correlated with a high or a low number of edges. Of course, more detailed properties of the graphs such as the number of connected components, the number of singletons, and the size of the largest component could also be important. 

\begin{figure*}[htb!]
\vspace{-0.15in}
\centering
\subfigure[]{
\includegraphics[width =.375 \textwidth]{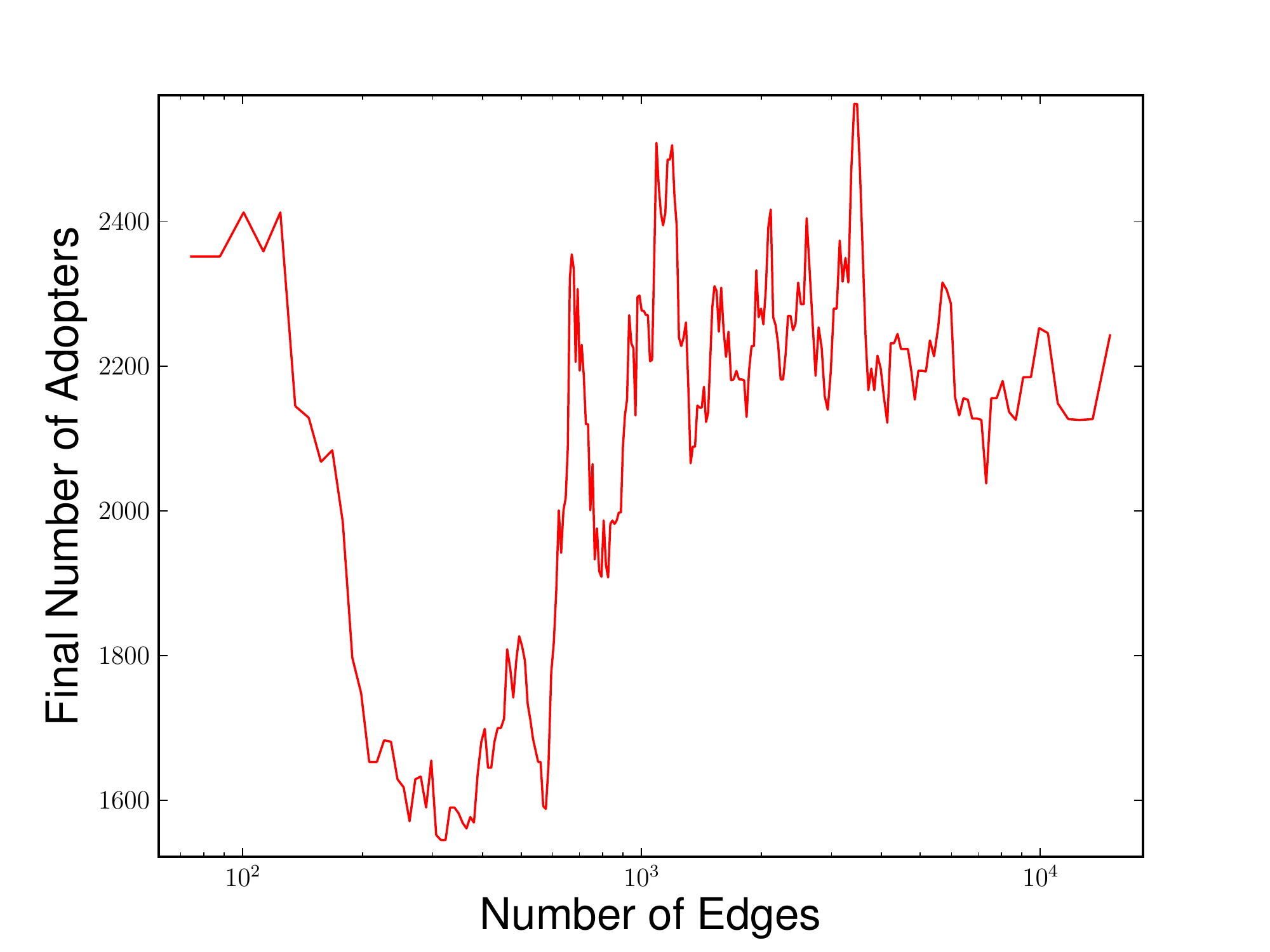} 
\label{f:edges_init_1000_final_adopters}
}
\vspace{-0.05in}
\subfigure[]{
\includegraphics[width =.375 \textwidth]{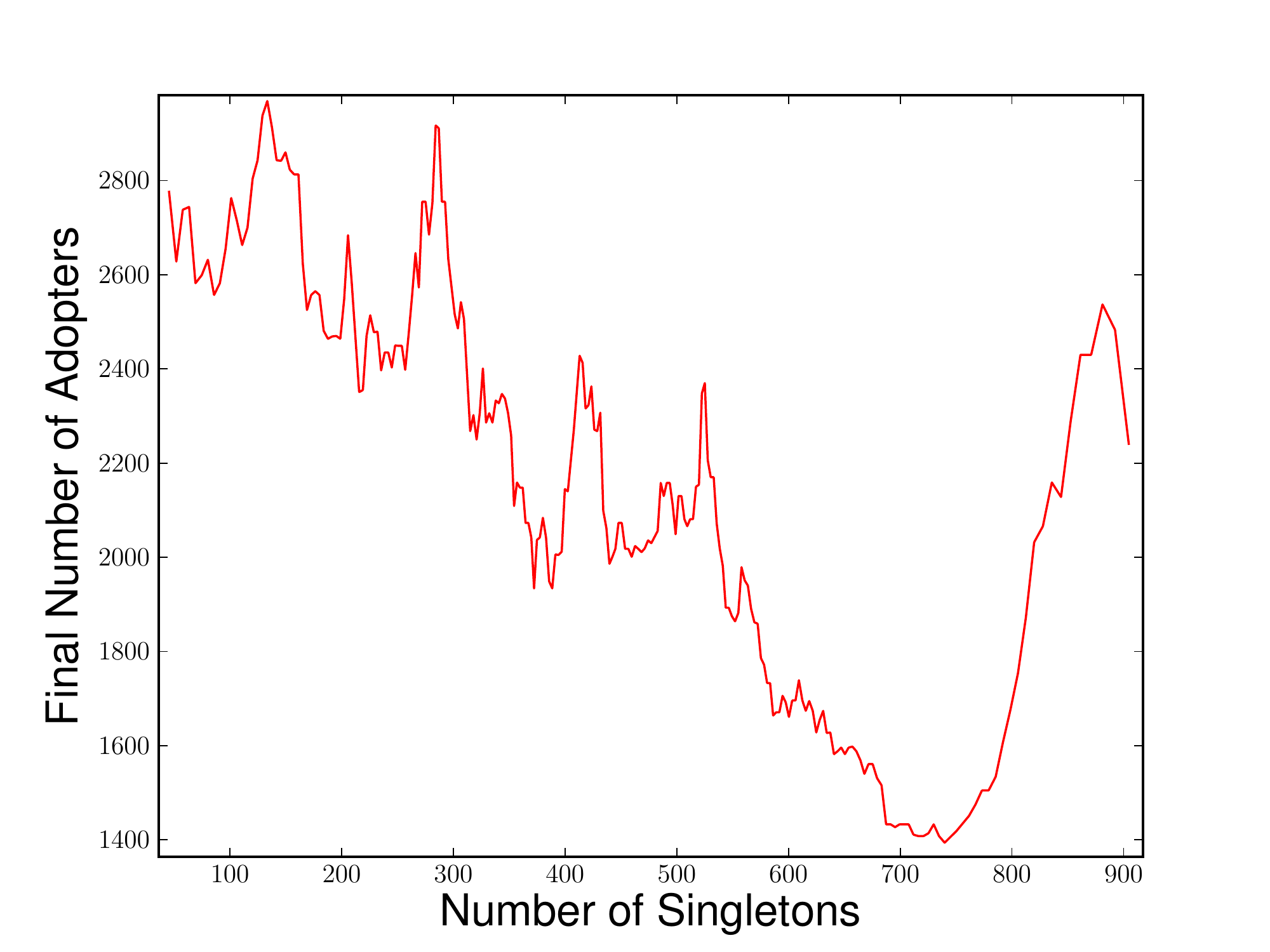} 
\label{f:singletons_init_1000_final_adopters}
}
\subfigure[]{
\includegraphics[width =.375 \textwidth]{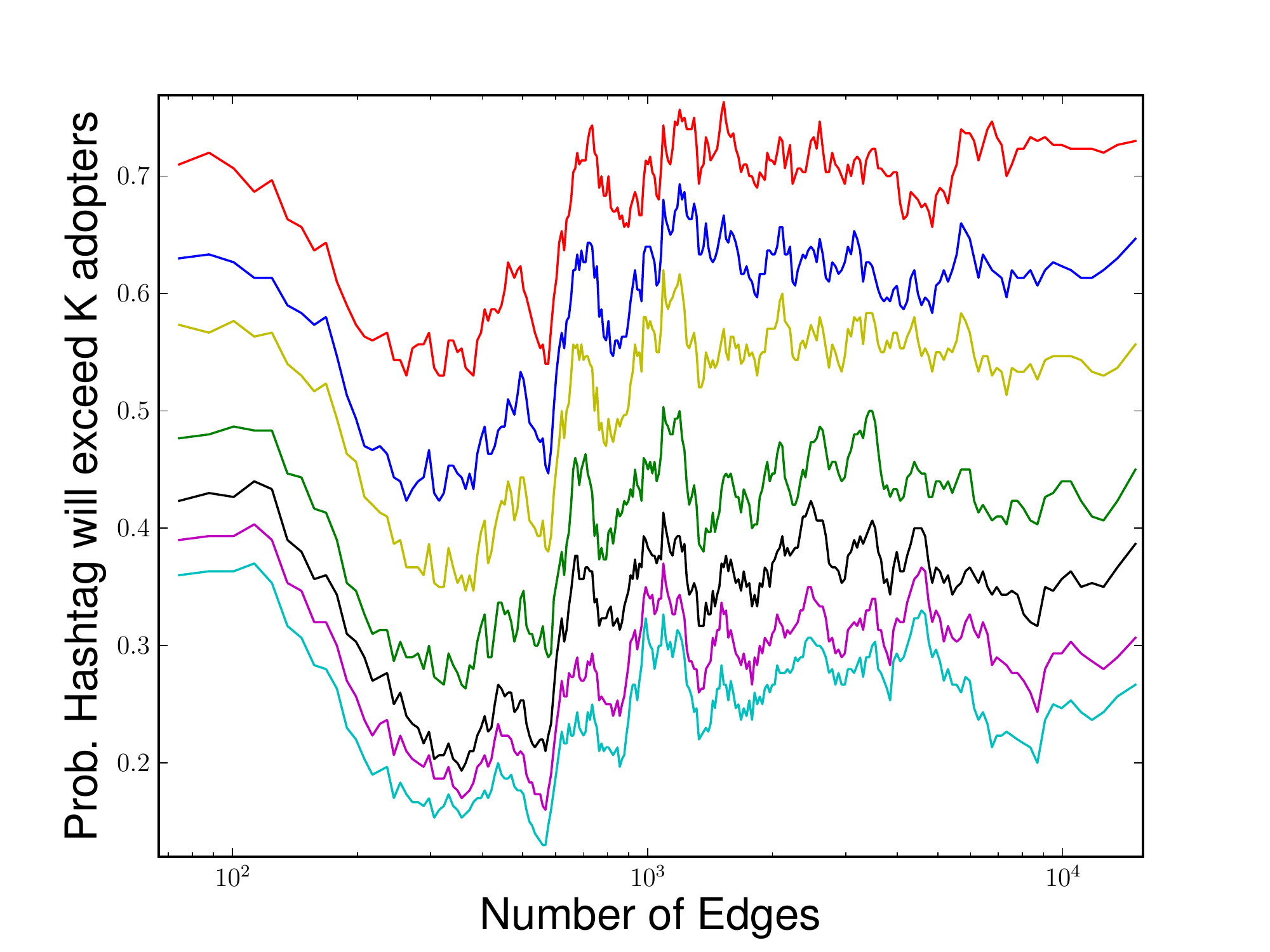} 
\label{f:edges_init_1000}
}
\subfigure[]{
\includegraphics[width =.375 \textwidth]{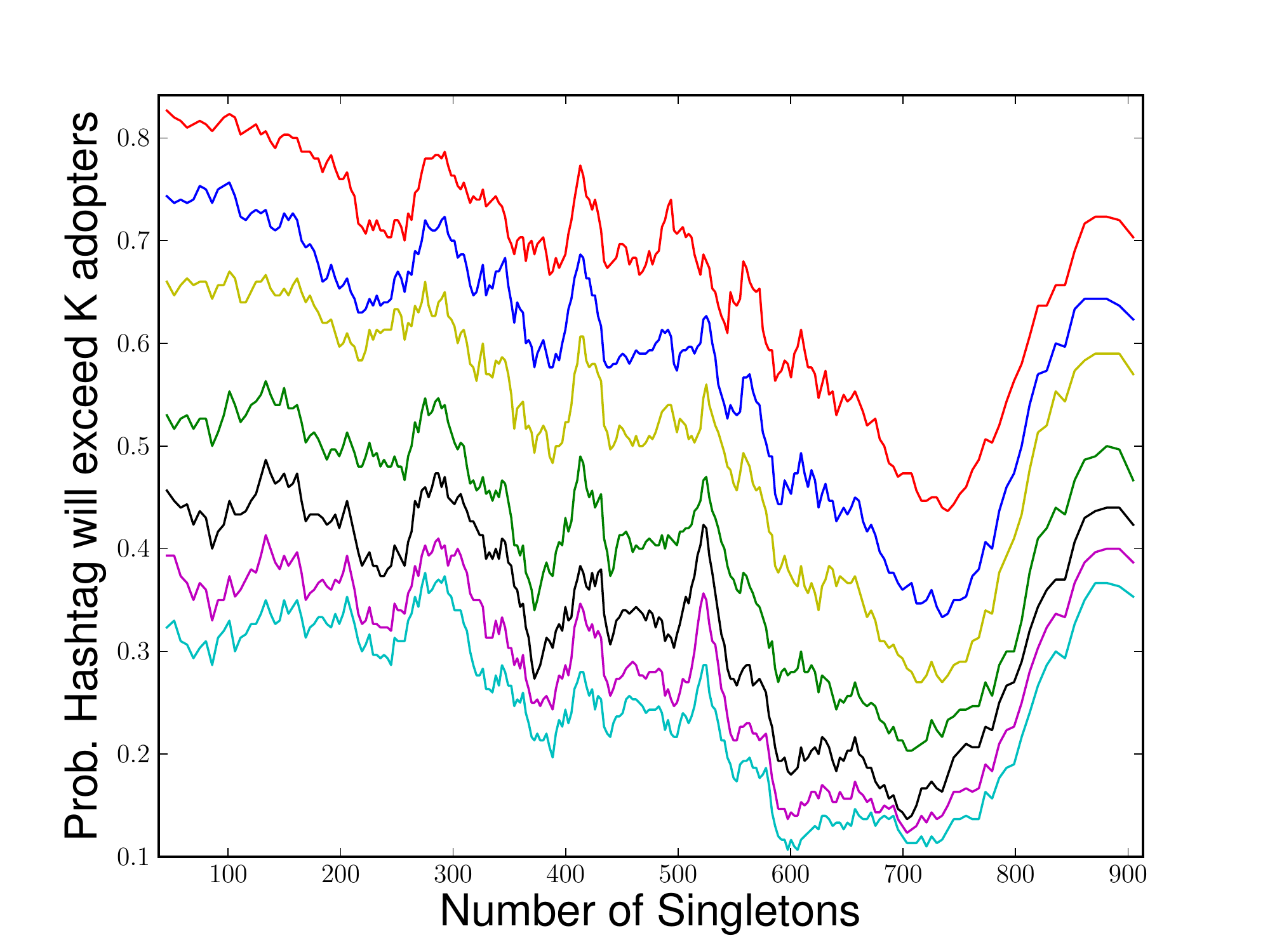} 
\label{f:singletons_init_1000}
}
\vspace{-0.1in}
\caption{Median number of final adopters as a function of the number of (a) edges and (b) singletons in the graph induced by the 1000 initial adopters, using a sliding window. Probability that hashtags will exceed $K$ adopters given the number of (c) edges and (d) singletons in the graph induced by the 1000 initial adopters, using a sliding window. From top to bottom, $K = 1500, 1750, 2000, 2500, 3000, 3500, 4000$. We observe that hashtags with many or few singletons or edges are more likely to grow than hashtags with intermediate amounts. \label{f:singletons_edges_init_1000_final_adopters}}
\vspace{-0.05in}
\end{figure*}

We begin by exploring how different structural properties of the initial graph induced by early adopters affect the probability that a hashtag will become popular. We consider all 7397 hashtags in our data that had at least 1000 adopters, and construct the follower graph induced by the first 1000 users. For each hashtag, we compute the number of users who eventually used the hashtag and the number of edges and singletons in their corresponding initial graphs. Figures \ref{f:edges_init_1000_final_adopters} and \ref{f:singletons_init_1000_final_adopters} show that the number of eventual adopters does not monotonically change with the number of singletons or edges, instead we find an interior minimum. This suggests that hashtags with either many or few edges and singletons tend to grow larger than hashtags with an intermediate number of singletons and edges. The corresponding graphs for different choices of number of initial adopters, besides 1000, produce similar curves. 

In practice,  one is not always interested in the exact final number of adopters in a diffusion process. Instead, it is desirable to know if the number of adopters will surpass a certain threshold or if the adopter population will double, triple, etc. In Figures \ref{f:edges_init_1000} and \ref{f:singletons_init_1000} we plot the probability that a hashtag with 1000 adopters reached $K=$ 1500, 1750, 2000, 2500, 3000, 3500 and 4000 adopters as a function of the number of singletons and edges in the subgraph of the initial 1000 adopters. Again, the likelihood of growth is highest when the singleton and edge counts are either very large or very small. Furthermore, the trends are consistent for the different choices of $K$, suggesting that the trend holds for short, medium, and long time frames. We conducted the same experiment with different numbers of initial adopters (the 2000 initial adopters and the 4000 initial adopters), and we observe the same results.

Many of the hashtags in our dataset correspond to popular world events. For example \emph{\#iranelection} was used during the disputed 2009 elections, and \emph{\#michaeljackson} was used after the death of Michael Jackson. Such hashtags tend to become very large and their initial set of users are not necessarily connected in Twitter. Hence, the hashtags with a small number of connections among initial users could correspond to hashtags about popular exogenous events where the first adopters are just the first ones to learn about the event. On the other hand, hashtags with very dense initial graphs (such as \emph{\#tcot} and \emph{\#tlot}, respectively for top conservatives on Twitter and top libertarians on Twitter) may correspond to very ``viral" topics. That is, once a user gets exposed to the hashtag after seeing that her friend used it, she is very likely to use it, making the hashtag very popular. The hashtags that are in the middle of these two extremes lack both virality and exogenous force and hence do not obtain many adopters. Interestingly, we find that these two competing effects generate an interior minimum that we can observe at large scale in our data. However, we note that these hypotheses are only meant as possible explanations and it is an open problem to design experiments that provide further evidence that these are the mechanism that explain the observed interior minimum.

\begin{figure*}[htb!]
%\begin{center}
\centering
\includegraphics[width =0.92 \textwidth,]{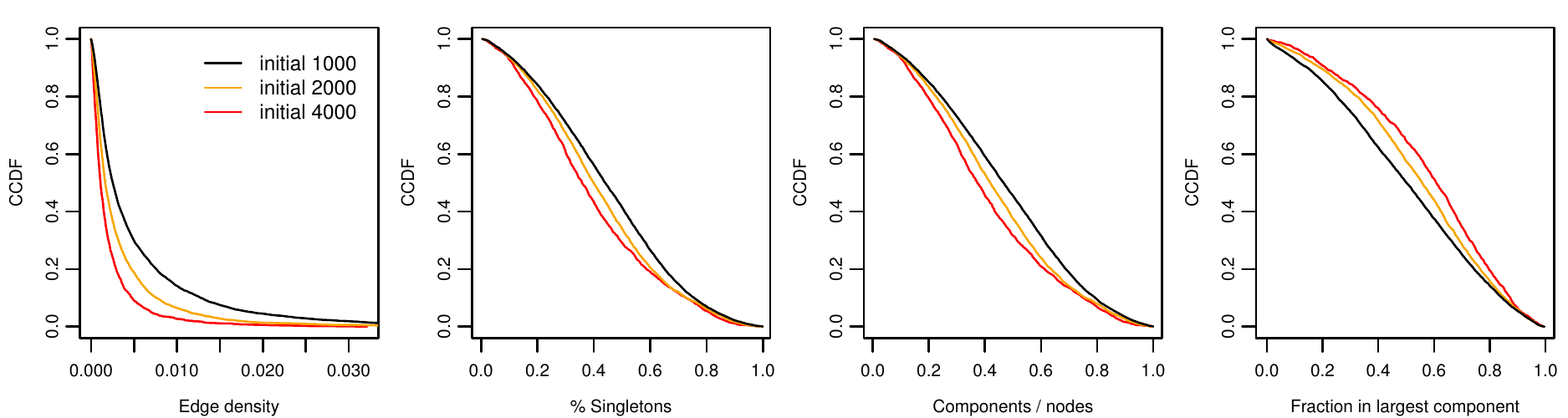} 
%\end{center}
\vspace{-0.05in}
\caption{Distribution of the structural features of the subgraphs induced by the 1000, 2000, and 4000 initial adopters. We see that while the edge count exhibits a heavy tailed distribution, the number of singletons, components and the size of the largest component are all broadly distributed over their support. \label{f:distro}}
%\vspace{-0.05in}
\end{figure*}

Having observed that the number of singletons and edges of the initial graph can be informative with respect to a hashtag's growth, we now ask whether these two features, and more generally the structure of the initial graph, can actually predict whether the hashtag will obtain many additional adopters. In order to select appropriate features for a prediction model, it is important to understand the different kinds of connections we can identify on Twitter based on follower and @-message relationships.\\

\vspace{-0.05in}

\noindent{\bf{Informational and social edges.}} In Twitter, users can unilaterally follow or @-message other users without having to ask for their approval. This environment allows for the connections of users to have different meanings. For example, if two users on Twitter are friends in real life, they are likely to follow each other. However, if a user is interested in another user, but they do not actually know each other, the following relation may be unreciprocated. We refer to this type of relationship as informational. For example, celebrities are followed by many of their fans, but they don't usually follow their fans back. Hence, we could think of Twitter as a network composed in two kinds of edges: social and informational \cite{Cheng:2011}. Given this lack of consistency of the meaning of connections on Twitter, we will now attempt to tease apart the two kinds of connections as they may have different prediction potential.

Formally, given the directed follower network on Twitter which we refer to as the \emph{full graph}, we define a directed edge from user $A$ to user $B$ as \emph{informational} if $A$ follows $B$ but $B$ does not follow $A$. We define an undirected edge between users $A$ and $B$ as \emph{social} if they follow each other. Next, we define the \emph{social graph} as the network of Twitter users and their social edges only, and the \emph{informational graph} as the users with their informational edges only. Note that the social graph is undirected, and the full and informational graphs are directed. Note that we can define corresponding graphs using @-message edges instead of the follower edges in a similar way. \\

\noindent{\bf{Predictive model.}} We train a logistic regression model to predict whether the number of adopters of a hashtag will eventually double. We examine other levels of growth further on. We use simple topological properties of the full graph, the social graph, and the informational graph. For each graph we compute the number of edges, number of singletons, number of connected components (weakly connected components for the full and informational graphs), and the size of the largest connected component. We train separate logistic regression models with the same features but based on the @-message and follower graphs.

\begin{table*}
\scriptsize
\vspace{-0.05in}
\centering
\begin{tabular}{c||c|c|c|c|c|c}
Model Features & F: 1k$\rightarrow$2k & F: 2k$\rightarrow$4k & F: 4k $\rightarrow$ 8k & @$\ge$3: 1k$\rightarrow$2k & @$\ge$3: 2k$\rightarrow$4k & @$\ge$3: 4k$\rightarrow$8k  \\ 
\hline
\hline
All  & \bf{0.674} & \bf{0.668} & \bf{0.683} & \bf{0.568} & 0.574 & 0.590 \\
Social Graph Only & 0.639 & 0.639 & 0.642 & 0.565 & 0.568 & 0.590 \\
Full Graph Only & 0.658 & 0.659 & 0.658 & 0.568 & 0.571 & 0.581 \\
Info. Graph Only & 0.642 & 0.649 & 0.663 & 0.559 & \bf 0.576 & \bf 0.601\\
\hline
\# Full Graph Edges  & 0.556 & 0.525 & 0.548 & 0.534 & 0.506 & 0.547 \\
\# Social Edges  & 0.538 & 0.529 & 0.545 & 0.530 & 0.510 & 0.547 \\
\# Conn. Comps, Full Graph & 0.579 & 0.591 & 0.603 & 0.525 & 0.511 & 0.535 \\
\# Conn. Comps, Social Graph & 0.567 & 0.575 & 0.593 & 0.527 & 0.504 & 0.533\\
\# Conn. Comps, Info. Graph & 0.566 & 0.565 & 0.585 & 0.528 & 0.504 & 0.546\\ 
\# Singletons, Full Graph & 0.583 & 0.599 & 0.602 & 0.529 & 0.513 & 0.548 \\
\# Singletons, Social Graph & 0.574 & 0.579 & 0.595 & 0.525 & 0.516 & 0.548\\
\# Singletons, Info. Graph & 0.571 & 0.566 & 0.590 & 0.525 & 0.513 & 0.545 \\
Max Comp. Size, Full Graph   & 0.573 & 0.587 & 0.599 & 0.521 & 0.520 & 0.541 \\
Max Comp. Size, Social Graph  & 0.555 & 0.581 & 0.598 & 0.520 & 0.520 & 0.537 \\
Max Comp. Size, Info. Graph & 0.556 & 0.565 & 0.586 & 0.527 & 0.504 & 0.547 \\
\hline
\hline
Majority Vote & 0.518 & 0.521 & 0.547 & 0.518 & 0.521 & 0.547 
\end{tabular}
\normalsize
\vspace{-0.05in}
\caption{Accuracy of a logistic regression mode for predicting whether a hashtag will double the number of adopters at different starting points: the 1000, 2000, and 4000 initial adopters, for both the follower and the @-message graphs. The accuracy of all models was evaluated using 10-fold cross-validation. \label{t:predict_growth}}
\end{table*}

To understand our ability to meaningfully separate graphs based on the structural features we analyze, in Figure~\ref{f:distro} we plot the complementary cumulative density functions for the features, as computed for all hashtags that exceeded size 1000, 2000, and 4000. We present the number of singletons, the number of components, and the number of adoptees in the largest component. The three features are broadly distributed across their support, consistently for all three subgraph sizes.

For each feature, we include its value as well as the logarithm of the value. Note, for example, that the distribution of edge densities suggests it is more useful to use the logarithm of this feature. Additionally, for every feature except for number of edges, we include a ``distance from the mid-point'' transformation: $|v_f - \frac{m_f}{2}|$, where $v_f$ is the value of the feature and $m_f$ is the largest possible value of the feature. The reason for this transformation is that, as figure \ref{f:singletons_edges_init_1000_final_adopters} suggests, high growth of the hashtags may be correlated with large or small values of some features. Having this transformation allows the algorithm to capture this trend. We do not include this transformation for number of edges because for the number of edges, $m_v$ is extremely large, and few of the hashtags we consider have an edge density greater than $0.5$,\footnote{We use edge density here since we have different initial sizes. But for a single prediction task, edge density is equivalent to the number of edges since the number of users is the same.} making all these transformed features linearly dependent upon the initial feature. 

We begin by using our logistic regression model to predict whether a hashtag will double in size. For each hashtag $h$ that has at least $k$ adopters, we compute the features of the model and predict whether $h$ will eventually obtain $2k$ adopters. We run the prediction task using the follower graph and the @-message graph separately. For the @-message graph we used a threshold of at least 3 @-messages to form an edge, having observed that different message count thresholds produce similar results. Table \ref{t:predict_growth} shows the accuracy of the full multivariate model using all the features and transformations as well as a model using each standalone feature and its transformations. We evaluated the accuracy using 10-fold cross-validation for $k=1000, 2000,$ and $4000$. Using the follower graph we obtain an accuracy of around 67\%. This is 14 percentage points above a baseline of around 53\% obtained from a naive majority vote algorithm, which simply classifies all hashtags as ``yes'' if the majority of hashtags doubled and ``no'' if the majority of hashtags do not. Using the @-message graph we do not perform as well as with the follower graph. For the @-graph, the accuracy of the full model is 57\% compared to a baseline of 53\%. Furthermore, we find that the accuracy changes very little for different choices of $k$, which suggests that classifying hashtags doubling does not get harder of easier we change the original size of the hashtag. We note that it has also been observed in \cite{Tan+al:11a} that predictions based on the follower graph perform better than predictions based on the @-message graph.

Lastly, we compare the performance under different sets of features, i.e., using the \emph{Social Graph Only}, the \emph{Full Graph Only} or the \emph{Informational Graph Only}. We show that using the \emph{Social Graph Only} features cannot provide as good performance as the other two feature sets. Thus, we observe that it may be that informational relationships play a stronger role in the spread of hashtags. Furthermore, the \emph{Informational Graph Only} performs marginally better than the social graph, and in some case it performs marginally better than the full model with all the features included. We note that no feature alone provides good performance, indicating that it is important to consider varied aspects of the initial graphs. It is an interesting open problem to investigate if informational edges always carry more predictive power than social edges, and if so, to determine possible explanations for it.\\
 
\noindent{\bf Longer prediction horizons.} Having found that the original size of the hashtag does not affect the accuracy of the algorithm when we predict whether the size will double eventually, we now investigate whether the accuracy changes as we change the horizon of prediction. That is, what happens to the accuracy if we try to predict whether the hashtag will grow by a factor of $p$ for $p \in (1,\infty)$? We expect that when $p$ is close to 1 the algorithm will not gain much accuracy above the baseline for two reasons. First, the outcome will be very sensitive to noise since we are asking whether the hashtag will obtain just a few additional adopters, so finding the few hashtags that do not surpass the threshold becomes difficult. Second, because most hashtags will surpass the threshold, even the naive majority vote classifier will have high accuracy, leaving little room for improvement. Similarly, when $p$ is very large, we expect that the structure of the graph will lose predictive power, as the graph is itself changing as additional users adopt the hashtag, and this evolution may change the nature of its growth. Also, since most hashtags will not surpass the threshold, the majority vote classifier will again have high accuracy, again leaving little room for improvement. 

\begin{figure*}[htb!]
%\begin{center}
\vspace{-0.05in}
\centering
\includegraphics[width =0.75 \textwidth,]{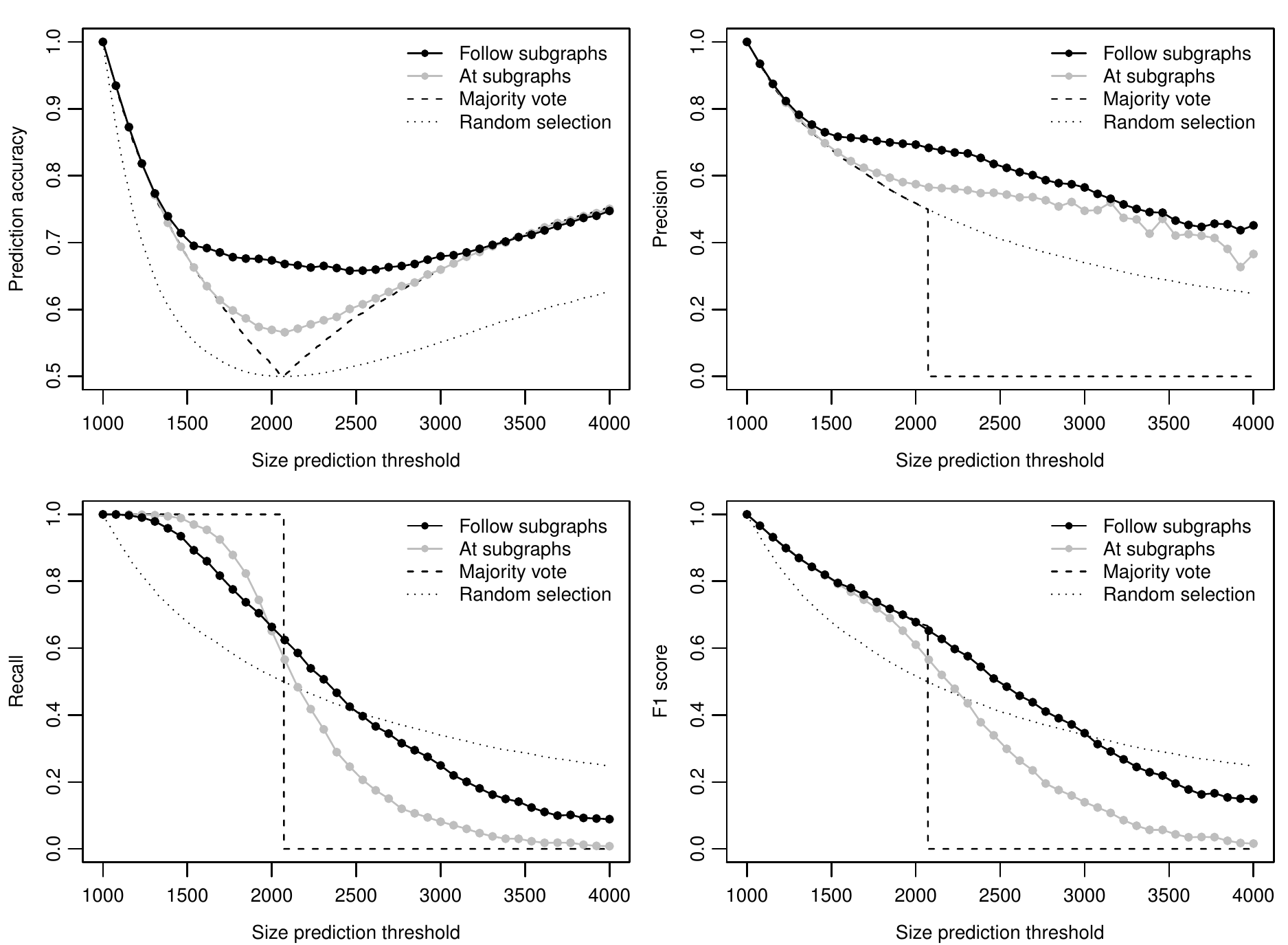} 
\vspace{-0.05in}
%\end{center}
\caption{Prediction accuracy, precision, recall, and F1 score when predicting wether a hashtag will exceed a certain size using our logistic regression model based on graph structure. Models were trained using 5-fold cross validation, applied to those 7397 hashtags that reached a size of 1000.}
\label{f:horizon}
\end{figure*}

To answer this question of accuracy at different horizons, we run logistic regression with the same features as above using the hashtags that had at least 1000 adopters and predicted whether they would reach at least $M$ users. Figure \ref{f:horizon} shows the accuracy, precision, recall, and F1 score of the full logistic model as a function of $M$. We compare these scores with two baseline naive classifiers -- the majority vote classifier discussed above, and a random algorithm that classifies as ``yes'' a random set of hashtags of size equal to the fraction of hashtags that obtained at least $M$ adopters.

We find that, indeed, the accuracy of our classifier is not above the baseline for very large and very small values of $M$. However, when we look at the precision of our classifier, we see that it stays above the baselines even for large values of $M$. That means that even for long term horizon prediction, the structure of the initial graphs maintains predictive power. The recall of the classifier starts off reasonably large for small values of $M$, but drops as $M$ gets very large. This is due to the fact that when $M$ is large, the number of hashtags that surpass $M$ adopters will be very small, making it very hard to identify all the few that did. 

In summary, our growth classifier maintains an accuracy of roughly 70\%. Its accuracy stays above our baselines for mid-term horizons and it equals the baselines for long and short term horizons. Its precision decreases slightly as the horizon increases, but it always stays above our baselines. Its recall starts out high, but it drops dramatically as the horizon increases. It is optimal for a classifier to have high precision and recall, and in our case we are able to maintain high precision, but recall falls for long term horizons. Depending on the actual application of the classifier, sometimes it may be more desirable to have high recall than precision and vice-versa. A possible application for wanting to know the future size of a certain hashtag is market forecasting. If people are using hashtags that correspond to new products, and they would like to use social media analysis to track which product to apply their advertising budget towards, having high precision is preferable over having good recall. On the other hand, low recall simply means that many products that became popular were not identified by the algorithm, so opportunities were lost, but not investments.

\section{Related Work}

There have been many studies on collaborative tagging systems and on Twitter.
But considerably less work has been done on the interplay between social and topical structure.

Collaborative tagging systems, which allow users to share their tags for particular resources, form the basis of our understanding of hashtags \cite{Golder:2006,Marlow:2006,Halpin:2007,Thom-Santelli:2008,Ramage:2009,Yin:2009,Hsu:2010,Schifanella:2010,Huang:2010}.  
Golder and Huberman \cite{Golder:2006} first studied the usage patterns of tags and developed a model for pattern prediction. 
Marlow et al. \cite{Marlow:2006} performed an early study on Flickr, developing a taxonomy of social tagging systems. They found that friends show a larger similarity in vocabulary compared with a random user baseline, which suggests that social links and tag usage is indeed related in other domains besides Twitter.  

Markines et al. \cite{Markines:2009} proposed a general and extensive foundation for the formulation of similarity measures in folksonomies, such as matching, overlap, mutual information, and Jaccard, Dice, and cosine similarity to study topics. 
A quite related study is a recent work by Schifanella et al. \cite{Schifanella:2010} on the interplay of the social and semantic components of social media on Flickr and Last.fm.  They showed that a substantial level of local lexical and topical alignment is observable among users who lie close to each other in the social network. Their analysis suggests that users with similar topical interests are more likely to be friends, and therefore semantic similarity measures among users based solely on their annotation metadata should be predictive of social links. The structure and the temporal evolution of social network has been investigated in several papers \cite{Kumar:2006,leskovec:www2008}. Leskovec et al. \cite{leskovec:www2008} studied local mechanisms driving the microscopic evolution of social networks. Furthermore, there has been work on the prediction of future connections in a social network based on the current ones \cite{Taskar03linkprediction,Liben-Nowell:2007,Schifanella:2010}. The work of
\cite{Leroy:2010,Rossetti:2011} also examined the problem of link prediction in different setting, e.g., in cold start and in multidimensional networks.
In contrast, our work uses simple features to show that the set structure of topics can help predict social relationships and that social structures within early adopters can be used to predict the eventual popularity of the topic. 
Other works have studied the tagging system from different perspectives. For example, 
Ramage el al. \cite{Ramage:2009} proposed a generative model based on Latent Dirichlet Allocation (LDA) that jointly models text and tags in such settings.

\section{Conclusion}

In this study, we find that the user-generated hashtag set system on Twitter and the topology of the connections among users are two fundamentally related structures. We show that this relation can be useful for predicting social links and predicting topic popularity using only simple features. By studying distance measures among users based on common hashtags, we find that we are able to predict, with reasonably high accuracy, the links between the users. Furthermore, the size of the smallest common hashtag turns out to be a very good predictor of linkage despite being one of cheapest ones to compute. 
We also find that combining the topical overlap of the users with the structural features of the graph induced by the shared topics can dramatically improve the accuracy of the prediction task. A possible applications of this would be any situation where one knows the topical interests of users and would like to predict connections among them, including recommendations systems for connections in online social networks. For example, recommending who to follow on Twitter based on an individual's hashtag usage, or recommending people to friend on Facebook based on an individual's ``likes.''

After observing how simple structural features of the graph induced by a hashtag are useful in predicting social connections, we discover that they are also useful for predicting the popularity of the hashtag itself. These features are very efficient to compute in $O(|V|+|E|)$ time. We find that the future popularity of the hashtags does not monotonically increase with the density of the graph induced by its initial set of users. Instead, we find that the popularity of the hashtag is highest when the density is either very low or very high. This is an interesting finding that offers a different perspective to the ideas from ``viral marketing'' where a small number of connections could be considered a negative property with respect to future growth.

Throughout our study we compare our results generated by the follower graph and the @-message graph. We find interesting distinctions among the two, notably that @-connections are easier to predict, but they turn out to be less informative when predicting the popularity of a hashtag from the connections of early adopters. Also, since the @-graph can be viewed as a weighted graph, we are able to compare our results at different tie strengths by considering graph with only @-edges with at least $k$ @-messages. We find that stronger ties are easier to predict, but they do not provide better or worse predictive power with regard to future hashtag popularity. \\

\noindent\textbf{Acknowledgments.} We thank Jon Kleinberg and Lillian Lee for useful discussions. Also, we thank Brendan Meeder for providing data. This work was supported in part by
NSF grants IIS-0910664 and IIS-1016099. 
Any opinions, findings, and conclusions or recommendations 
expressed in this material are those of the authors and do not necessarily 
reflect those of the National Science Foundation.

%\small
\fontsize{9pt}{10pt} \selectfont
\bibliographystyle{aaai}
\bibliography{references}

\end{document}